\documentclass[useAMS,usenatbib]{mn2e}
\usepackage{amsmath,fleqn,graphicx,xcolor,amssymb}
\usepackage{multirow}
\renewcommand{\[}{\begin{equation}}
\renewcommand{\]}{\end{equation}}
\def\p{\partial}

\def\agamaTwo{{\sc agama}b}
\def\agama{{\sc agama}}

\let\boldgrk=\gkvecten
\let\boldgrksc=\gkvecseven

\def\gkthing#1{{\mathchoice%
	{\hbox{{\boldgrk\char#1}}}
	{\hbox{{\boldgrk\char#1}}}
	{\hbox{{\boldgrksc\char#1}}}
	{\hbox{{\boldgrksc\char#1}}}}}

\def\vtheta{\gkthing{18}}

{\newif\ifnotend
\notendtrue
\def\veclist{ABCDEFGHIJKLMNOPQRSTUVWXYZabcdefghijklmnopqrstuvwxyz.}
\def\top#1#2.{#1}
\def\tail#1#2.{#2.}
\loop\expandafter\xdef\csname v\expandafter\top\veclist\endcsname%
{{\noexpand\bf\expandafter\top\veclist}}
\edef\veclist{\expandafter\tail\veclist}
\if\veclist.\notendfalse\fi\ifnotend\repeat}
{\newif\ifnotend
\notendtrue
\def\veclist{ABCDEFGHIJKLMNOPQRSTUVWXYZ.}
\def\top#1#2.{#1}
\def\tail#1#2.{#2.}
\loop\expandafter\xdef\csname c\expandafter\top\veclist\endcsname%
{{\noexpand\cal\expandafter\top\veclist}}
\edef\veclist{\expandafter\tail\veclist}
\if\veclist.\notendfalse\fi\ifnotend\repeat}
\def\eqrf#1{(\ref{#1})}
\def\d{{\rm d}}

\def\Lc{L_{\rm circ}}

\def\Jcrit{J_{z\rm crit}}
\def\Jrcrit{J_{r\rm crit}}

\def\rd{}
\def\e{\mathrm{e}}

\def\fracj#1#2{{\textstyle{#1\over#2}}}

\title[Actions of highly eccentric orbits]
{Actions of highly eccentric orbits}

\author[Thomas J Wright \& James Binney]{
  Tom Wright$^1$ James Binney$^2$\thanks{E-mail: binney@physics.ox.ac.uk}\\  
  $^1$Somerville College, Oxford OX1\\
  $^2$Rudolf Peierls Centre for Theoretical Physics, Clarendon Laboratory,
  Oxford, OX1 3PU, UK\\
}

\begin{document}
\maketitle

\begin{abstract}
The challenge presented by computing actions for eccentric orbits in
axisymmetric potentials is discussed. In the limit of vanishing angular
momentum about the potential's symmetry axis, there is a clean distinction
between box and loop orbits. We show that this distinction persists into the
regime of non-zero angular momentum. In the case of a St\"ackel potential,
there is a critical value $I_{3\rm crit}(E)$ of the third integral $I_3$ below
which $I_3$ does not contribute to the centrifugal barrier. An orbit is of
box or loop type according as its value of $I_3$ is smaller or greater than
$I_{3\rm crit}$. We give algorithms for determining $I_{3\rm crit}(E)$ and the
critical action $\Jcrit$ below which orbits in any given potential are
boxes.  It is hard to compute the actions and especially the frequencies of
orbits that have $J_z\simeq\Jcrit$ using the St\"ackel Fudge. A modification
of the Fudge that alleviates the problem is described.
 \end{abstract}


\section{Introduction}

The Sun and most of its neighbours are on orbits that are not highly
eccentric. By contrast, most of the stars that comprise the Galaxy's stellar
halo are on highly eccentric orbits, and simulations of the clustering of
dark matter suggest that most dark-matter particles are also on highly
eccentric orbits. It follows that highly eccentric orbits will be important for
any realistic Galaxy model.

Angle-action variables were invented to analyse the dynamics of the solar
system, and our understanding of the structure and evolution of  planetary
systems is hugely dependent on insights obtained from angle-action
coordinates. In light of this fact,  it's natural to seek angle-action
coordinates for galaxies.

Highly eccentric orbits are not especially difficult to compute by numerical
integration of the equations of motion, but such integrations do not deliver
values for an orbit's radial and vertical actions, $J_r$ and $J_z$. Knowledge
of orbits' values of these constants of motion is valuable because actions
have many properties that other constants of motion, such as the orbits'
initial conditions do not possess. Moreover, actions are the only constants
of motion that can be embedded as the `momenta' of a system of canonical
coordinates, the conjugate variables being the `angle' coordinates
$\theta_i$.

Galaxies are not precisely in equilibrium but they are nearly in equilibrium,
so their dynamics and evolution are naturally understood via perturbation
theory. Action-angle coordinates are the bedrock on which Hamiltonian
perturbation theory rests.

To be useful, angle-action coordinates must be computable. Classically they
are derived from the solution $S(\vx,\vJ)$ of Hamilton-Jacobi equation --
$S(\vx,\vJ)$ is the generating function of the canonical transformation
from ordinary phase-space coordinates $(\vx,\vv)$ to angle-action coordinates
$(\vtheta,\vJ)$.  Unfortunately, this nonlinear partial differential equation
on real space can be solved only for very special potentials, and we know
(from the occurrence of orbit trapping at resonances) that it is in principle
impossible to solve it for the actual potentials of galaxies.  Consequently,
in the case of a real galaxy we must be content with systems of angle-action
coordinates such that the actions are only approximately constant. Then we
can solve for the time evolution of the actions by perturbation theory. 

The PhD thesis of P.T.\ de Zeeuw  \citep{deZeeuw1985} was seminal because it
established that galaxies have gravitational potentials that are in
significant respects similar to the potentials that St\"ackel investigated a
century earlier, which give rise to separable Hamilton-Jacobi
equations.

Currently three schemes for assigning suitable angle-action coordinates to
phase-space points $(\vx,\vv)$ are available: the `St\"ackel Fudge'
\citep{JJB12:Stackel} yields $(\vtheta,\vJ)$ given $(\vx,\vv)$;
\cite{SaJJB14} give an algorithm for determining the actions of numerically
integrated orbits; and torus
mapping \citep{JJBPJM16,BinneyVasilievWright} yields $(\vx,\vv)$ given
$(\vtheta,\vJ)$. In this paper we focus on difficulties that arise when these
techniques are applied to highly eccentric orbits. 

Section \ref{sec:orbits} reviews the nature of orbits in axisymmetric
potentials and explains how the sharp division of these orbits into two
classes {\rd(boxes and loops)} at $J_\phi=0$ persists to non-zero $J_\phi$.  It explains the role
that $J_z$ plays in establishing a centrifugal barrier at small radii and
explains how the action $\Jcrit$ that divides the box and loop regimes can be
computed. Section \ref{sec:Fudge} explains how the St\"ackel Fudge can
perform poorly when $J_z\simeq\Jcrit$ and $J_z$ is contributing only
marginally to the barrier. A workaround is proposed in
Section~\ref{sec:solution}. Section~\ref{sec:conclude} sums up and explains
why the algorithm given in Section~\ref{sec:getJcrit} is important.

\section{Orbits in a flattened potential}\label{sec:orbits}

Orbits with $J_z=0$ are confined to the equatorial plane. Each orbit occupies
an annulus whose width increases with $J_r/J_\phi$ -- circular orbits have
$J_r=0$. The value of $J_z/J_\phi$ indicates an orbit's inclination $i$, which
is zero when $J_z=0$ and tends to $90\,$deg as $J_z/J_\phi\to\infty$.  An
orbit's eccentricity is quantified by $J_r/L$, where $L=J_z+|J_\phi|$ is the
extension to a non-spherical potential of the total angular momentum.  Orbits
with $J_r=0$ are circular if $J_z=0$, or `shell orbits' when $J_z>0$ -- these
are analogues of the inclined circular orbits of a spherical potential. 

An instructive starting point for the study of eccentric orbits is the
action-space plane $J_\phi=0$.  The orbits of this plane are confined in real
space to a plane whose axes we call $x$ and $z$. The $z$ coordinate is the
same as the galaxy's vertical coordinate while $x$ runs along a line
within the equatorial plane. Within this plane, the gravitational potential
is barred, $x$ being the long axis and $z$ being the short axis.
Consequently, orbits with $J_\phi=0$ divide into loop orbits and box orbits
\citep[e.g.][\S3.3]{GDII}.
The simplest box orbit comprises an oscillation along the $x$ axis. A general
box orbit is obtained by adding an oscillation parallel to the $z$ axis. The
actions $J_x$ and $J_z$ quantify the amplitudes of these two oscillations.
When $J_z$ exceeds a critical value, the $x$ and $z$ oscillations lock
together to form a loop orbit, which occupies an elliptical annulus in real
space. Now $J_z$ quantifies the central semi-major axis of the annulus and $J_r$
quantifies the annulus's width: $J_z$ has morphed into a generalised angular
momentum and $J_r$ has become a measure of eccentricity.

\begin{figure}
\centerline{\includegraphics[width=.8\hsize]{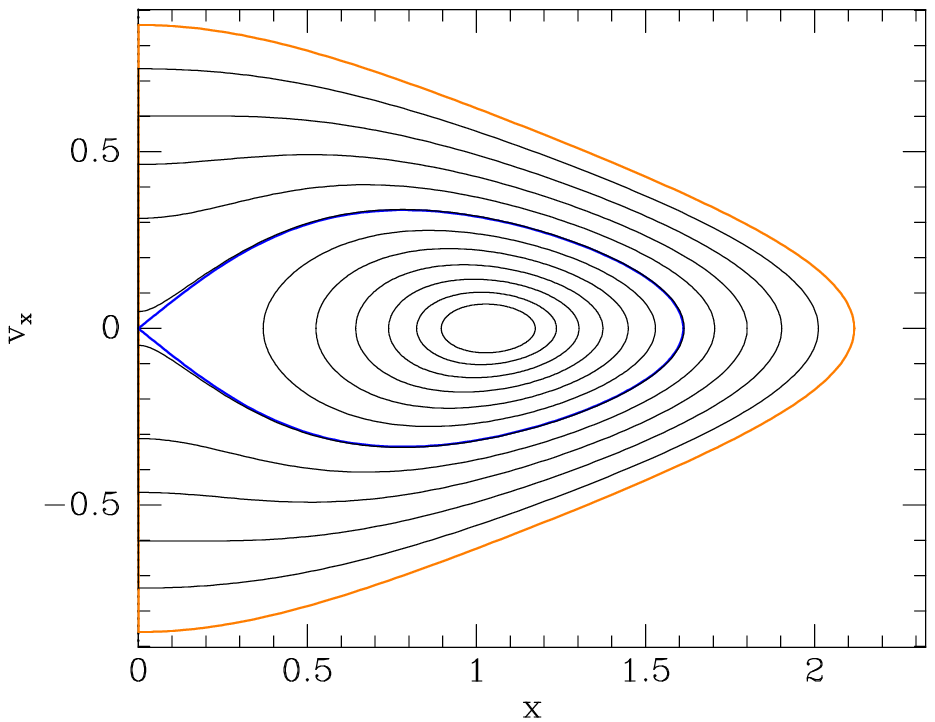}}
\caption{Surface of section at $E=\Phi(0)/2$ and $J_\phi=0$ in the potential
of a perfect oblate ellipsoid with unit mass, {\rd scale length $a=1$}, and axis
ratio $q=0.6$. At this
energy the circular angular momentum is $\Lc=0.663$.}\label{fig:StackSoS}
\end{figure}

\begin{figure}
\centerline{\includegraphics[width=.8\hsize]{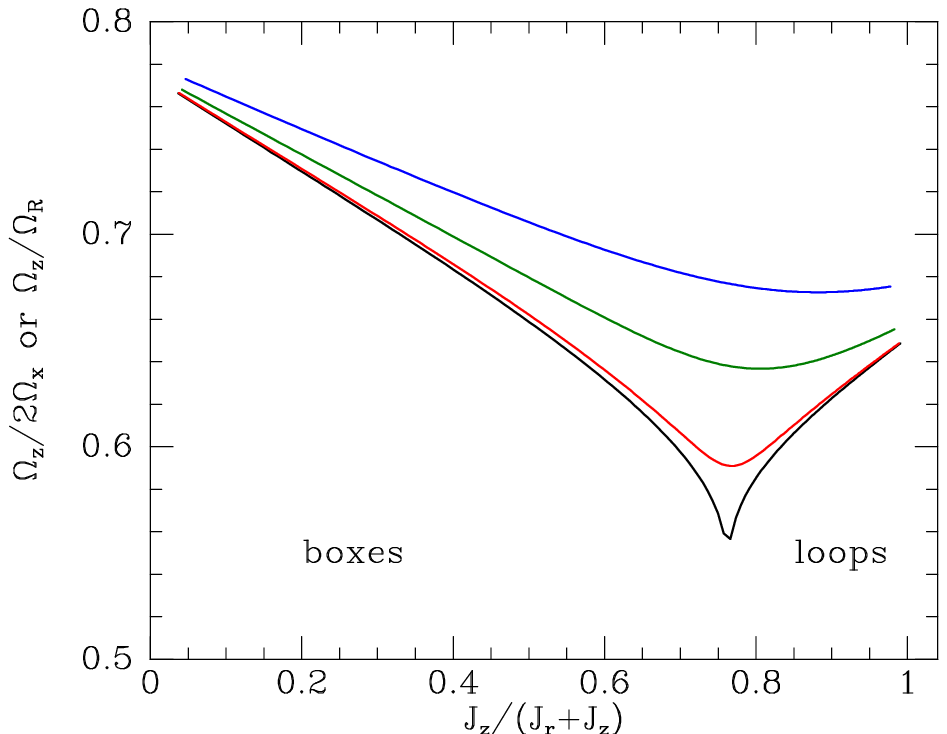}}
\caption{The ratio of vertical to horizontal frequencies as a function of
circularity $J_z/(J_r+J_z)$  for the orbits with the energy of
Fig.~\ref{fig:StackSoS}. The four
curves are for $J_\phi=0$ (black), $J_\phi=0.02$ (red), $J_\phi=0.1$ (green) and $J_\phi=0.2$
(blue).  We plot $\Omega_z/2\Omega_x$ for box orbits 
and $\Omega_z/\Omega_R$ for loop orbits.}
\label{fig:StackSoSFreq}
\end{figure}

Fig.~\ref{fig:StackSoS} is a typical $(x,v_x)$ surface of section (SoS) for orbits
with $J_\phi=0$ -- the figure is computed for a perfect ellipsoid
\citep{deZeeuw1985} of unit mass and
scale-length $a$, and axis ratio $q=0.6$, but the corresponding surface of
section for many galactic potentials would be qualitatively the same. Each
curve is a cross section through an orbital torus $\vJ=\hbox{constant}$. All
tori have the same energy, so the value of $J_r$ follows once $J_z$ is
specified. {\rd The  orange curve is the torus $J_z=0$ and $J_z$ increases
monotonically as we proceed inwards. The curves that lie between the orange
and blue curves are generated by box orbits, while those that lie inside the
blue curve are generated by loop orbits.} At the centre of the smallest
curve lies the point generated by the shell orbit $J_r=0$. Let $\Jcrit$ be
the value of $J_z$ for the blue curve that separates boxes from
loops.

The black curve in Fig.~\ref{fig:StackSoSFreq} shows the variation with the
circularity
$c\equiv J_z/(J_r+J_z)$ of the ratio of the vertical and horizontal frequencies at
the energy of Fig.~\ref{fig:StackSoS}.  The curve's sharp minimum occurs at
$J_z=\Jcrit$. Box orbits lie to the left of this minimum, loop orbits lie to
its right.  In the regime of box orbits, the frequency $\Omega_z$ of vertical
oscillations exceeds the long-axis frequency $\Omega_x$. As $J_z$ increases,
the two frequencies converge. At the transition, the horizontal frequency
switches from $\Omega_x$ to $\Omega_r\simeq2\Omega_x$ and in the loop-orbit
regime $\Omega_z/\Omega_r$ rises with $J_z$.

\begin{figure}
\centerline{\includegraphics[width=.8\hsize]{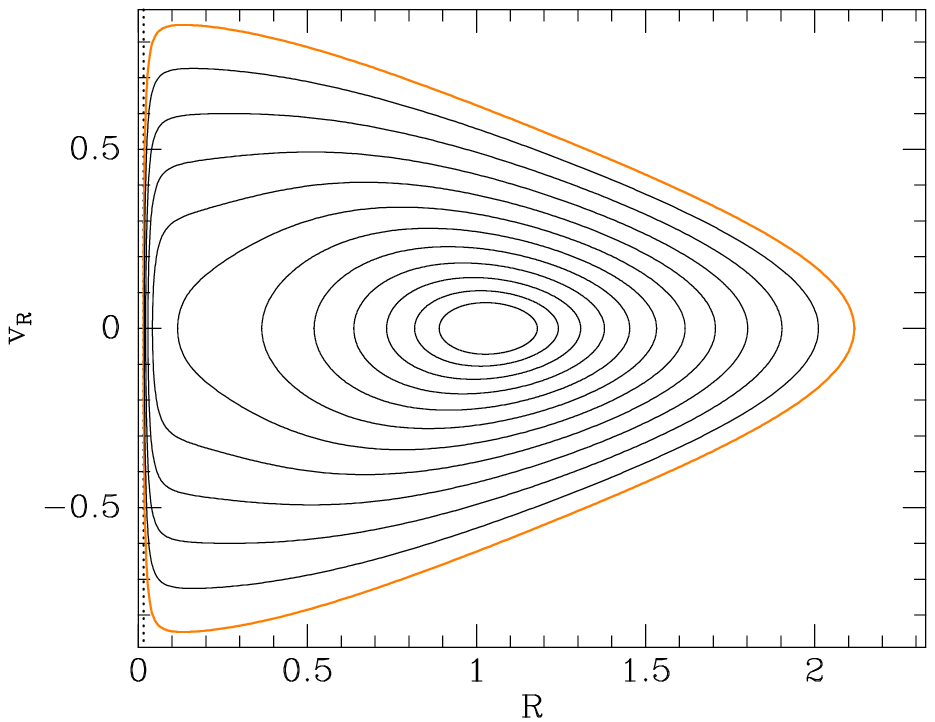}}
\caption{Surface of section at $J_\phi=0.02$ for the orbits with the energy of
Fig.~\ref{fig:StackSoS}.}\label{fig:StackSoSL02}
\end{figure}

Fig.~\ref{fig:StackSoSL02} is the surface of section for the same potential
and energy as  Fig.~\ref{fig:StackSoS} but $J_\phi=0.02$ -- for
comparison, the circular angular momentum at the plot's energy is
$\Lc=0.633$. Now the orange curve $J_z=0$ does not cross the $v_R$
axis but plunges almost vertically very close to the axis, as do sections of
four of the black curves within it.  The smaller curves in
Fig.~\ref{fig:StackSoSL02} don't have near-vertical left-hand sections. {\rd The
analogue of the blue curve in Fig.~\ref{fig:StackSoS} would lie between the smallest
black curve that has a vertical segment and the largest black curve that
doesn't.}

The
red curve of Fig.~\ref{fig:StackSoSFreq} shows the corresponding variation of
the frequency ratio, which differs from that when $J_\phi=0$ mainly in the
disappearance of the cusp at $J_z=\Jcrit$. In particular, the red curve divides
into falling and rising segments either side of $\Jcrit$, so the distinction
between box and loop orbits has not been lost by introducing non-zero
$J_\phi$ even though the appearance of the SoS has changed qualitatively
between Figs.~\ref{fig:StackSoS} and \ref{fig:StackSoSL02}. The green and
blue
curves show the frequency ratios when $J_\phi=0.1$ and $0.2$, respectively.
The curve for $J_\phi=0.1$ slopes upwards on the extreme right, while when
$J_\phi=0.2$ the upward-sloping section is on the point of vanishing. 

Physically orbits with infinitesimal $J_\phi$ differ infinitesimally from
orbits with vanishing $J_\phi$ notwithstanding the qualitative change in the
SoS between Figs.~\ref{fig:StackSoS} and \ref{fig:StackSoSL02}: when
$J_\phi=0$ the orbit repeatedly cuts the $z$ axis, while when
$J_\phi=\epsilon$ it passes the axis at $y\simeq\epsilon/v_x$; when
$J_\phi\ne0$ the orbit is not confined to the $xz$ plane but moves a bit
further from this plane with each near-miss of the axis. Whereas with
$J_\phi=0$, the sign of $v_R$ instantaneously changes at $x=0$, with
$J_\phi\ne0$ the reversal of $v_R$ occurs continuously over the small range in
$x$ at which $R=\sqrt{x^2+y^2}$ is not dominated by $x$. Hence it is natural
that the distinction between box and loop orbits, which is evident at
$J_\phi=0$, is also significant when $J_\phi$ is non-zero but small.

\begin{figure}
\centerline{\includegraphics[width=.8\hsize]{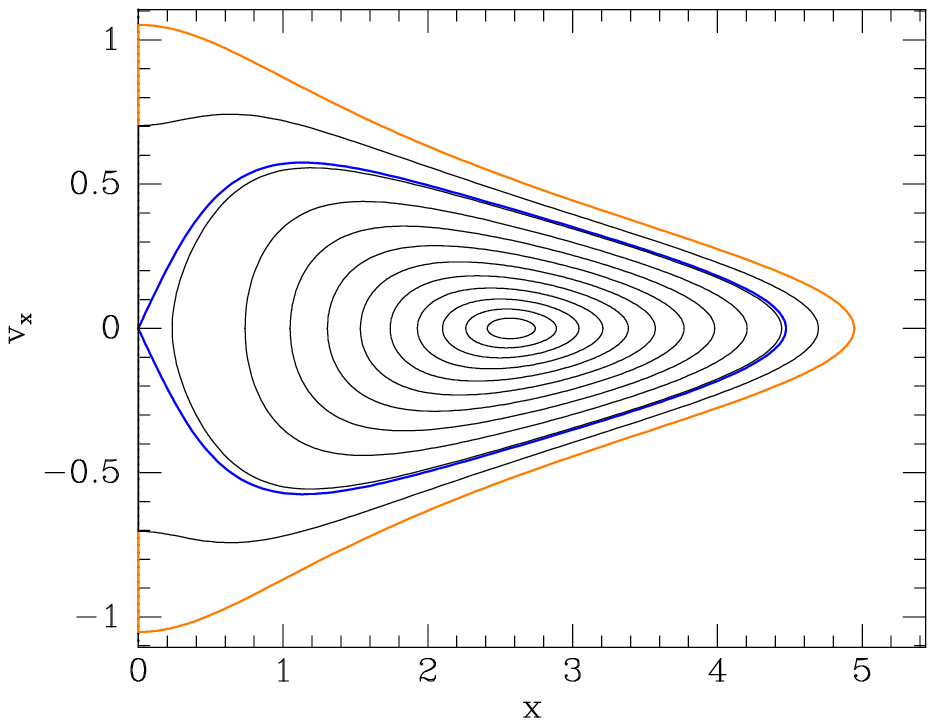}}
\vskip5pt
\centerline{\includegraphics[width=.8\hsize]{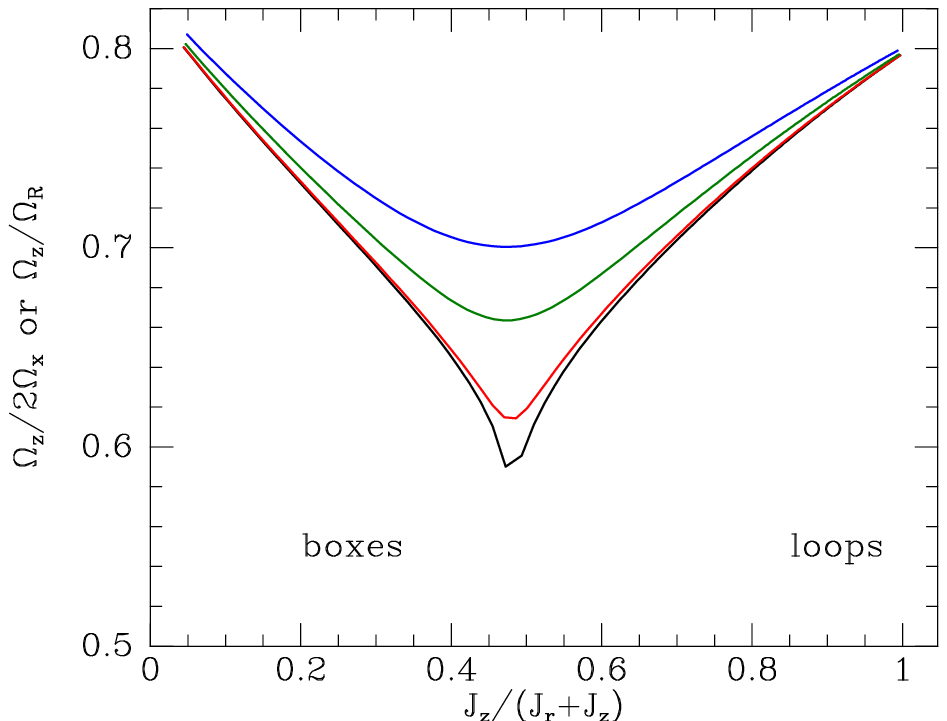}}
\caption{The analogues of Figs.~\ref{fig:StackSoS} and \ref{fig:StackSoSFreq}
at a higher energy, namely $E=\Phi(0)/4$.}\label{fig:StackSoShigherE}
\end{figure}

Fig.~\ref{fig:StackSoShigherE} shows analogues of Figs.~\ref{fig:StackSoS}
and \ref{fig:StackSoSFreq} at a higher energy, $E=\Phi(0)/4$. The curves
extend to larger $J_z$ because more energy is available, but $\Jcrit$, which
lies at the minimum of the black curve, has not increased so much, with the
consequence that the right-hand segments of the curves, associated with loop
orbits, have grown in importance relative to the left-hand segments, which
are associated with box orbits. As $J_\phi$ increases, the minimum becomes
less sharp but moves very little horizontally. The largest possible value of
$J_z$ decreases as $J_\phi$ increases because $J_\phi$ is being varied at
constant energy and it ties up energy that cannot be invested in $J_z$. It
follows that the value of $J_z$ at the minimum is a decreasing function of
$J_\phi$.

\begin{figure}
\centerline{\includegraphics[width=.9\hsize]{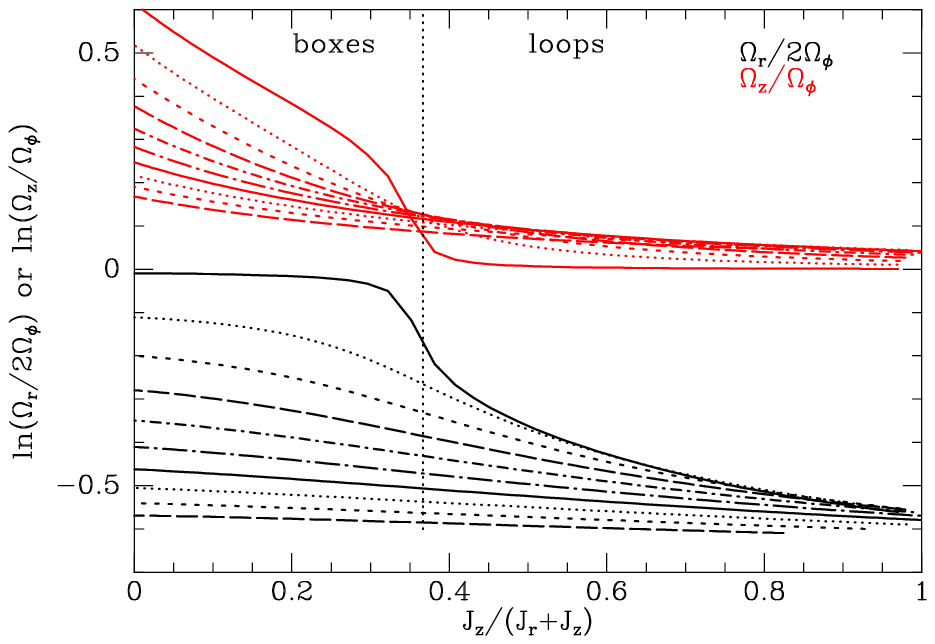}}
\caption{Frequency ratios for orbits of low $J_\phi$ in a flattened perfect
ellipsoid. All orbits pass through $(R,z)=(3a,0)$, where $a$ is the scale
radius of the ellipsoid. In the case of the top full curve in each panel the velocity there is
$(v_R,v_z,v_\phi)=(V_m\sin\theta,V_m\cos\theta,0.01V_c)$ where  $V_c$ is the
circular speed at $R$ and $V_m$ yields the energy of the circular orbit at
$R$.  The $n^{\rm th}$ broken curves below the top curve on the left
corresponds to $V_\phi=(0.01+n/10)V_c$.}\label{fig:staeckel2a}
\end{figure}

Fig.~\ref{fig:staeckel2a} shows frequency ratios that involve $\Omega_\phi$
as functions of circularity at several fixed values of $J_\phi$. The full
black curve, which is for $J_\phi$ equal to one percent of its circular
value, shows that $\Omega_r=2\Omega_\phi$ in the box-orbit regime on the
left, while $\Omega_r$ falls more and more below $2\Omega_\phi$ in the
loop-orbit regime on the right. The full red curve shows that $\Omega_z$
exceeds $\Omega_\phi$ in the box-orbit regime but equals it in the loop-orbit
regime.  The broken curves generated by orbits with larger values of $J_\phi$
show that as $J_\phi$ is increased, the links between $\Omega_\phi$ and
$\Omega_r$ or $\Omega_z$ in the box-orbit or loop-orbit regimes become weaker
and the transition between the regimes becomes less sharp.

\subsection{Centrifugal barriers}\label{sec:centrifuge}

At the extreme left of Fig.~\ref{fig:StackSoSL02} there is a dotted vertical
line. This marks the pericentric radius of the planar orbit $J_z=0$, which is
set by $J_\phi$: it marks the `centrifugal barrier' imposed by non-vanishing
$J_\phi$. Box orbits are kept away from the $z$ axis by this barrier. Loop
orbits do not approach this barrier because they are subject to a centrifugal
barrier to which both $J_z$ and $J_\phi$ contribute. The simplest way to
grasp this phenomenon is to recall that the effective potential for motion in
a spherical potential has a contribution $L^2/2r^2$ where $L\equiv
J_z+|J_\phi|$ is the total angular momentum, so $J_z$ helps set the barrier.
When the potential is flattened, the contribution of $J_z$ to the
centrifugal barrier first becomes significant when $J_z$ reaches the critical
value $\Jcrit$. 

\begin{figure}
\centerline{\includegraphics[width=.8\hsize]{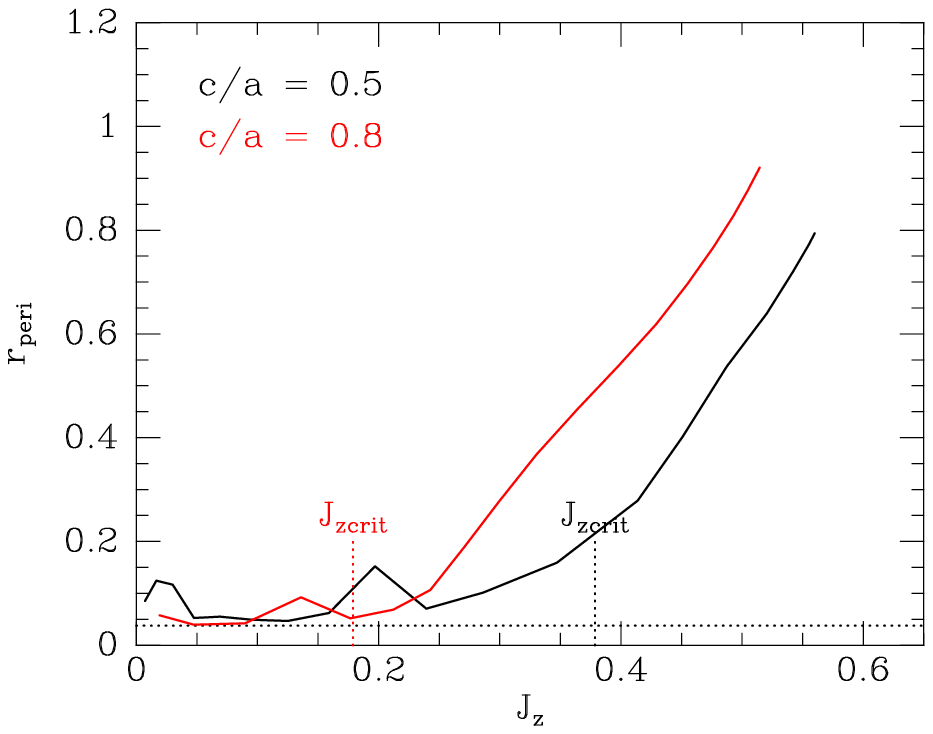}}
 \caption{{\rd The full lines show the smallest value of spherical radius
$r$} reached on numerically integrated orbits of energy $E=-0.3$ and
$J_\phi=0.05L_{\rm circ}(R_{\rm shell})$ in two NFW-like potentials. These
are  created by the density distributions of equation (\ref{eq:PhiNFW}) with axis ratio
$q=0.5$ (black) or $0.8$ (red). The critical values of $J_z$ in these
potentials are marked by vertical dotted lines. {\rd The dotted horizontal line
shows the smallest values of $r$} reached on the planar orbit defined by the
same values of $J_r$ and $J_\phi$ but $J_z=0$. The Fudge was used to compute
$J_z$.}\label{fig:getIcrit}
\end{figure}

\begin{figure}
\centerline{\includegraphics[width=.8\hsize]{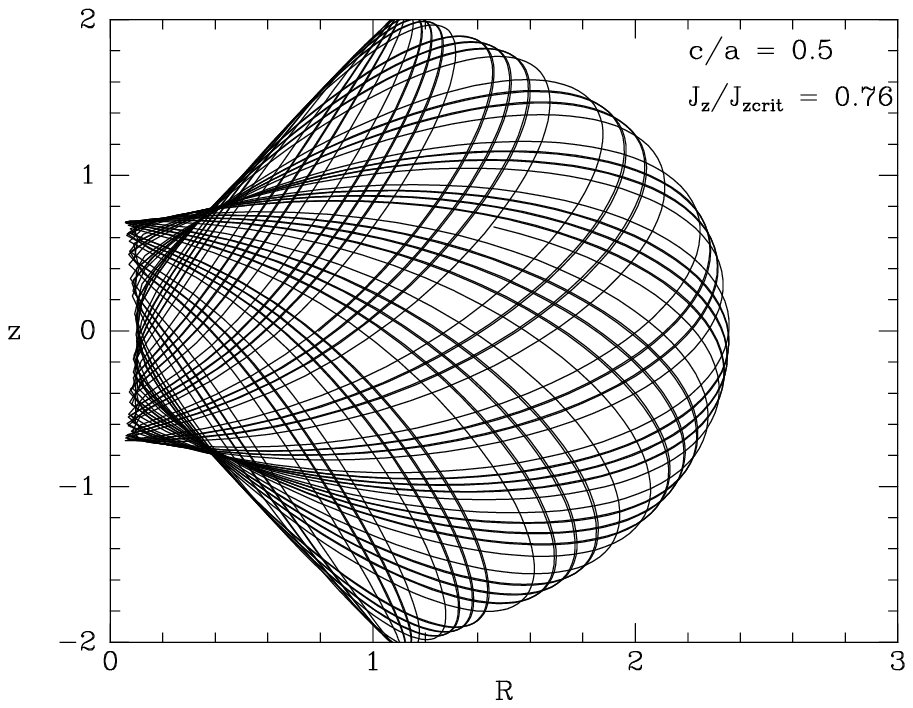}}
\centerline{\includegraphics[width=.8\hsize]{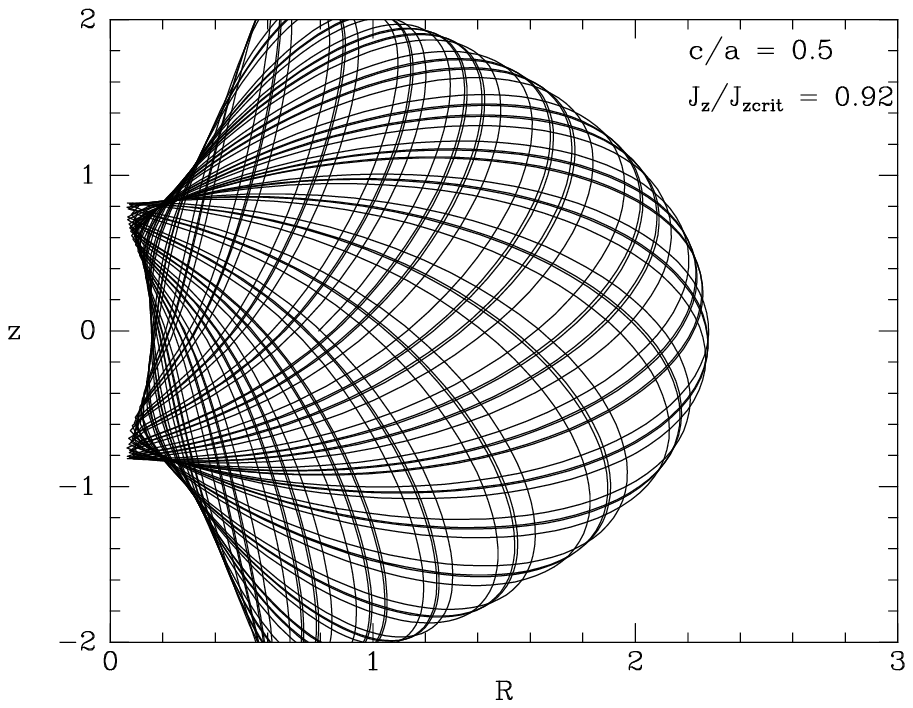}}
\caption{{\rd Orbits contributing to Fig.~\ref{fig:getIcrit} for $q=0.5$
either side of the box/loop transition.}}\label{fig:boxLoopBdy}
\end{figure}

{\rd The full curves in Fig.~\ref{fig:getIcrit} illustrates this phenomenon by
plotting the smallest distances from the centre, $r_{\rm peri}$ of orbits
with $J_\phi=0.05L_{\rm circ}$ in the potentials generated by bodies with the
\cite{NFW97} (NFW) density profile 
\[\label{eq:PhiNFW}
\rho(R,z)={1\over r(1+r)^2}\e^{-r^2/r_{\rm cut}^2},
\]
where $r^2={R^2+z^2/q^2}$ and $r_{\rm cut}=10$.
The potentials have axis ratio
$q=0.5$ (black curve) or to $q=0.8$ (red curve). As $J_z$ decreases, the
curves first plunge rapidly and then flatten off markedly when $J_z$ is
comparable to $\Jcrit$. Thereafter they asymptote to the pericentric radius
of the planar orbit $J_z=0$ with the same value of $J_\phi$, which is marked
by the horizontal dotted black line.  Wiggles above this line are caused by
centro-phobic resonances.}

{\rd Fig.~\ref{fig:boxLoopBdy} shows two orbits from Fig.~\ref{fig:getIcrit} that
might be considered to lie either side of the box/loop transition: the left
boundary of the upper orbit is nearly straight while that of the lower orbit
curves noticeably away from the origin. The orbits' values of $J_z/J_{z\rm
crit}$ are given at top right: even the loop-like lower orbit has
$J_z/J_{z\rm crit}<1$, implying that when $|J_\phi|>0$, loop orbits are reached
at lower values of $J_z$ than when $J_\phi=0$. In this context it's worth
noting that the loop-like orbit has $(J_z+J_\phi)/J_{z\rm crit}=1.016$.
}

\subsection{Determination of $\Jcrit$}\label{sec:getJcrit}

The last section established that $\Jcrit$ plays a significant role and it's
important to be able to determine it as a function of energy in realistic
potentials. This is a non-trivial task because {\rd in our experience} all
available algorithms for determining actions (the St\"ackel Fudge, torus
mapping and extraction of $S(\vx,\vJ)$ from a numerically integrated orbit)
fail when confronted with orbits near the box-loop transition. Here we
describe the algorithm for the determination of $\Jcrit(E)$ that we have
coded into the \agamaTwo\ \citep{BinneyVasilievWright} branch of the {\sc
agama} software library \citep{AGAMA}.

If we drop a star from a point $z=Z$ on the $z$ axis, the star will remain on
the axis and return after completing the (unstable) `short-axis orbit' of energy
$E=\Phi(0,Z,0)$. 
A box orbit with energy $E$ and $J_z\simeq\Jcrit(E)$ can be generated by
dropping a star from infinitesimally off the $z$ axis where $\Phi=E$. The
star falls to the centre, comes to rest at negative $z$ before returning and
coming to rest extremely close to its point of departure. From the orbit's
time sequence we can evaluate\footnote{$J_{\rm fast}$ would be the conserved
`fast' action in a perturbative treatment of the box-loop transition.}
\[
J_{\rm fast}={1\over2\pi}\oint \d z\,p_z.
\]
We now show that
\[\label{eq:JtotRes}
J_{\rm fast}=J_z+2J_r. 
\] 

We consider the case that the star is moving in a St\"ackel potential. Such a
potential is associated with a system $(u,v)$ of confocal ellipsoidal
coordinates such that
\begin{align}\label{eq:defUV}
R&=\Delta\sinh u\sin v\cr
z&=\Delta\cosh u\cos v,
\end{align}
 where $\Delta$ is the $z$ coordinate of one focus.  The system's ellipses in
the $xz$ plane are labelled by $u\ge0$ and its hyperbolae are labelled by $v$.
We exploit that
\[
p_R\d R+p_z\d z=p_u\d u+p_v\d v
\]
because both coordinate systems are canonical.
 
As the star falls from its point of release at $z=Z\leftrightarrow u=U$, to
the focus of the potential's coordinate system at $z=\Delta\leftrightarrow
u=0$, its value of $v$ is constant at $v=\pi/2$. Hence,
\[\label{eq:firstStep}
\int_Z^\Delta\d z\,p_z=\int_U^0\d u\,p_u.
\]
As the star travels from $z=\Delta$ to $z=-\Delta$, $v$ changes from $\pi/2$
to $-\pi/2$ while $u$ remains fixed at zero. So
\[
\int_\Delta^{-\Delta}\d z\,p_z=\int_{\pi/2}^{-\pi/2}\d v\,p_v.
\]
As the star moves from $z=-\Delta$ to $z=-Z$, $v$ is constant at $-\pi/2$
while $u$ returns from zero to $U$, so
\[
\int_{-\Delta}^{-Z}\d z\,p_z=\int_0^U\d u\,p_u,
\]
where $p_u>0$ whereas in equation (\ref{eq:firstStep}) $p_u<0$.
These three stages are then executed in reverse order, with reversed limits
of integration and changed signs on the momenta. Adding all the integrals, we
find
\[
J_{\rm fast}={1\over\pi}\oint \d u\,p_u+{1\over2\pi}\oint\d v\,p_v.
\]
This equation establishes equation (\ref{eq:JtotRes}) in the case of a St\"ackel
potential because then $2\pi J_r=\oint\d u\,p_u$ and $2\pi J_z=\oint\d
v\,p_v$. 

\begin{figure}
\centerline{\includegraphics[width=.9\hsize]{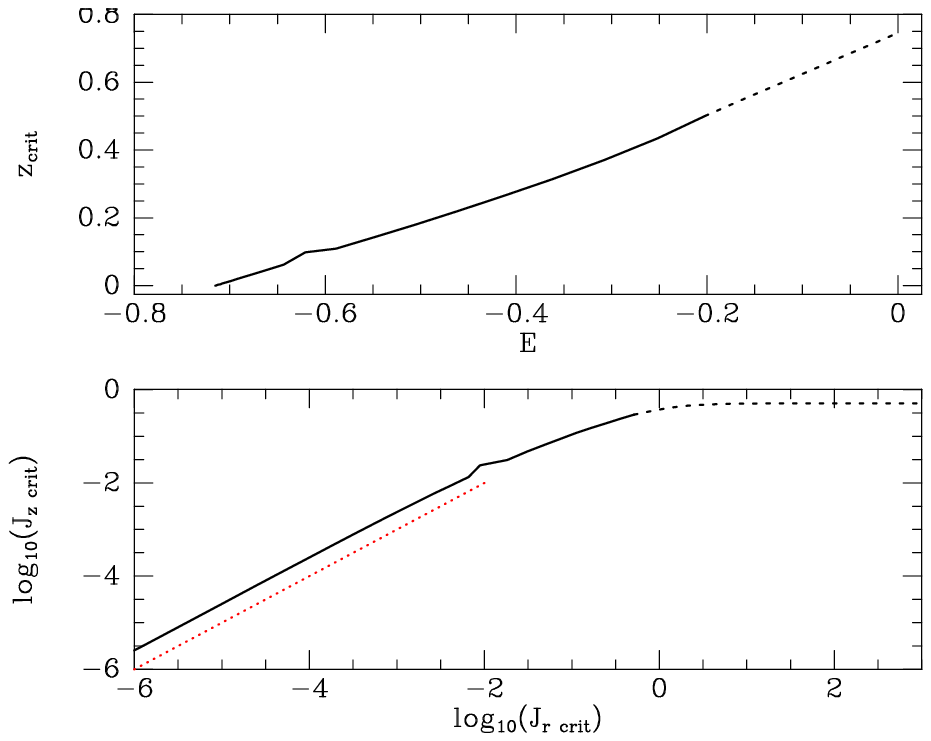}}
 \caption{Upper panel: $z_{\rm crit}$ versus $E$ for the NFW-like potential
of Fig.~\ref{fig:getIcrit} for axis ratio $q=0.8$. The full portion was
obtained from equation (\ref{eq:zcrit}) using values of $\Jcrit$ that follow
from equation (\ref{eq:JtotRes}) with $J_r$ estimated from a SoS.  The dashed
portion is the adopted linear extrapolation of the full portion. Lower panel:
the resulting graph of $\Jcrit$ as a function of $\Jrcrit$. The red dotted
line shows $y=x$.  }\label{fig:zcrit}
\end{figure}

Equation (\ref{eq:JtotRes}) generalises to any potential because $J_{\rm
fast}$ is defined by a closed path around an orbital torus so it has to be an
integer linear combination of the actions. The integer multipliers are
constant as we deform the potential into a St\"ackel potential.

The critical orbit's value of $J_r$, {\rd hereafter $\Jrcrit$,} can be obtained from the area inside the
orbit's trace in a SoS
such as Fig.~\ref{fig:StackSoS}. Substituting this value into equation
(\ref{eq:JtotRes}), we obtain $\Jcrit$.

As $E\to0$ (marginally bound orbits), it becomes hard to accumulate sufficient consequents for accurate
evaluation of the area so we employ the following extrapolation procedure. We
define $z_{\rm crit}(E)$ by
\[\label{eq:zcrit}
\Jcrit={2\over\pi}\int_0^{z_{\rm crit}}\d z\,p_z,
\]
 where $p_z(z|E)$ is the momentum along the short-axis orbit.  The argument
given above shows that in the case of a St\"ackel potential $z_{\rm crit} =
\Delta$ when $Z<\Delta$, where $Z$ is, as before, such that $E=\Phi(0,Z,0)$.
In the general case, we can solve equation (\ref{eq:zcrit}) for $z_{\rm
crit}(E)$ on a grid of values of $E$ low enough for $\Jcrit$ to be obtained
as described above.  At the largest of these $E$ values, $z_{\rm crit}$ tends
to a linear function of $E$ -- illustrated by the upper panel of
Fig.~\ref{fig:zcrit}. By extrapolating this function, we can predict $z_{\rm
crit}$ for higher $E$, and then obtain the corresponding values of $\Jcrit$
from equation (\ref{eq:zcrit}) -- the lower panel of Fig.~\ref{fig:zcrit}
shows the resulting function $\Jcrit(\Jrcrit)$ for one of the NFW-like
potentials of Fig.~\ref{fig:getIcrit}. 

At low energies
$\Jcrit\propto\Jrcrit$. {\rd Consequently, the line $(\Jrcrit,\Jcrit)$ divides
the portion of $(J_r,J_z)$ plane that lies below the line $H=E$, which runs from
top left to bottom right, into an upper triangle of loop orbits and a
lower triangle of box orbits such that the ratio of their areas, and thus the
relative sizes of the loop and box populations, are independent of $E$. At
high energies, by contrast, $\Jcrit$ asymptotes to a constant value, implying
that an ever increasing fraction of orbits are loops.}

\subsection{Numerical implementation}

When a potential is `created' in the software package \agamaTwo, box/loop
transition orbits are computed over essentially all energies so that
subsequently $\Jcrit$ can be recovered as functions of $J_{\rm fast}$ from
the potential's method {\tt getJzcrit}. Also stored at this stage are the
values $I_{3\rm crit}(E)$ of the transition orbits. They are evaluated at
$(\vx,\vv)=(R_{\rm shell},0,0,v_R,v_z,0)$, where $v_R$ is obtained from the
surface of section from which $J_r$ is computed and $v_z$ then follows from
the value of $E$ -- it matters where $I_3$ is evaluated because $I_3$ isn't
an exact constant of motion in a general potential. $I_{3\rm crit}(E)$ can subsequently be recovered from the
potential's method {\tt getI3crit}.
 
\section{Application to the St\"ackel Fudge}\label{sec:Fudge}

\begin{figure}
\centerline{\includegraphics[width=.8\hsize]{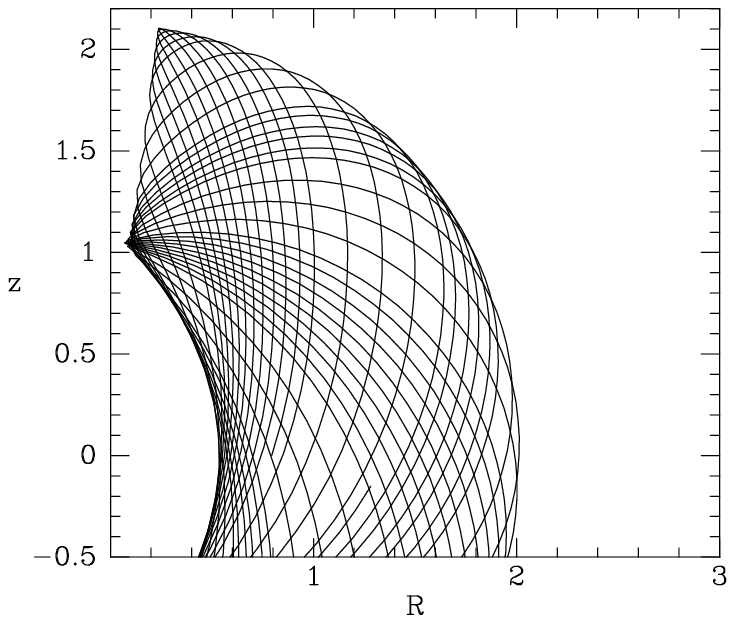}}
\vskip5pt
\centerline{\includegraphics[width=.85\hsize]{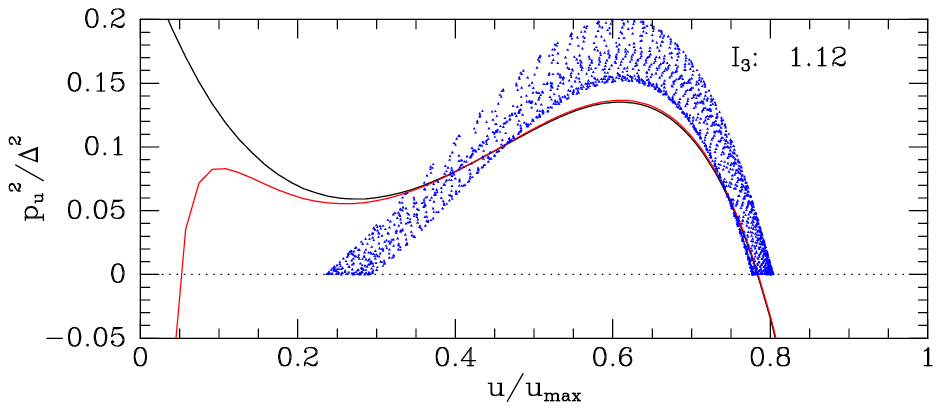}}
\caption{An orbit with $J_\phi=0.05L_{\rm circ}(R_{\rm shell})$ in the
NFW-like  potential used to make the
black curve in Fig.~\ref{fig:getIcrit}. The upper panel shows the orbit in the
meridional plane. The blue points in the lower panel show the
points $(u,p_u^2)$ through which the orbit passes when the focal distance of
the $(u,v)$ coordinate
system is fitted to the shell orbit at the given values of $E$ and $J_\phi$. The
red curve shows the values of $p_u^2(u)$ predicted by the full effective potential
(eqn.~\ref{eq:defI3}),
while the black curve shows the value of $p_u^2$ predicted by the effective
potential obtained by setting $J_\phi=0$.}\label{fig:problem}
\end{figure}

The St\"ackel Fudge has been widely used to compute actions in realistic
galactic potentials
\citep[e.g.][]{JJB12:dfs,SaJJB13b,Bovy2013,Bovy2014:streams,
BinneyWong,BinneySchoenrich2018,
Vasiliev2019,JBH_Galah2019,JBHThor2020}. It involves choosing a length
$\Delta$ that defines a coordinate system $(u,v)$ via equations (\ref{eq:defUV}), and
then applying to the given potential formulae that are strictly only valid for
a St\"ackel potential. In the widely used implementation of the Fudge
contained in the {\sc agama} code library \citep{AGAMA} $\Delta$ is set to the
value $\Delta_{\rm shell}$ that fits the shell obit $J_r=0$ that has the same
energy and angular momentum $J_\phi$ as the orbit under study.

The merit of a value of $\Delta$ is conveniently assessed by
computing an orbit in the given potential and then plotting the resulting data points
$(\vx_i,\vv_i)$ in the $(u,p_u)$ plane:
ideally $\Delta$ is chosen such that every point $(u,p_u)$ lies on a
curve rather than filling out a zone of non-zero area.

\subsection{A problem with the Fudge}

The upper panel of Fig.~\ref{fig:problem} shows the trace in the meridional
plane of an orbit in the potential associated with the black curve in
Fig.~\ref{fig:getIcrit}. The blue points in the lower panel show the
locations $(u,p_u)$ through which the orbit passes in the coordinate system
defined by the focal distance $\Delta_{\rm shell}$ that fits the shell orbit
of the given $(E,J_\phi)$. The red curve shows the values $p_u^2(u)$ upon
which the St\"ackel Fudge relies. This functional form arises from a
definition of the third integral $I_3$ of a St\"ackel
potential:\footnote{Eqn.~(\ref{eq:defI3}) is $E$ plus the definition of $I_3$ used by
\cite{JJB12:Stackel}. The definition used in both \agama\ and \agamaTwo\ differs by a factor
$\Delta^2$, so $I_3$ in the code has the dimensions of  action squared rather
than  energy. Discussion of changes in $\Delta$ is simpler without this
factor of $\Delta^2$.}
\[\label{eq:defI3}
I_3=E\cosh^2u-\frac{1}{2\Delta^2}\Big(p_u^2+{J_\phi^2\over\sinh^2u}\Big)-U(u).
\]
where $U(u)$ is the function that defines a St\"ackel potential through
\[\label{eq:defSPhi}
\Phi(u,v)={U(u)-V(v)\over\sinh^2 u+\sin^2 v}.
\]
The central idea of the St\"ackel Fudge is that for potentials $\Phi(u,v)$
similar to those of axisymmetric galaxies, a value $u_0$ can be chosen such
that
\[
\delta U\equiv(\sinh^2u+\sin^2v)\Phi(u,v)-(\sinh^2u_0+\sin^2v)\Phi(u_0,v)
\]
 has only weak dependence of $v$, so $\delta U+U(u_0)$ can be used in
equation \eqrf{eq:defI3} in place of $U(u)$. Equation \eqrf{eq:defI3} yields
$p_u^2(u)$ once the constant of motion $I_{3u}\equiv I_3+U(u_0)$ has been
determined from the initial conditions.  Similarly, $p_v^2$ can be determined
from 
\[\label{eq:defI3v}
I_{3v}=-E\sin^2v+\frac{1}{2\Delta^2}\Big(p_v^2+{J_\phi^2\over\sin^2v}\Big)-\delta
V(v),
\]
where $I_{3v}\equiv I_3-V(\pi/2)$ can be determined from the initial
conditions and we neglect the $u$ dependence of
\[
\delta V\equiv\cosh^2u\Phi(u,\pi/2)-\big(\sinh^2u+\sin^2v\big)\Phi(u,v).
\]

The
optimum value of $u_0$ is such that 
\[\label{eq:Rsh_uzero}
R_{\rm shell}=\Delta\sinh u_0,
\]
where $R_{\rm shell}$ is the radius at which the
shell orbit crosses the equatorial plane. Hence $u_0$ and $I_{3u}$ follow once
$\Delta$ has been chosen.

In Fig.~\ref{fig:problem} we plot $p_u^2/\Delta^2$ because $p_u$ is equal
to $\Delta$ times a linear combination of $p_R$ and $p_z$.  Consequently,
when we change $\Delta$, $p_u^2/\Delta^2$ does not change much at given $u$.
For this reason, equations (\ref{eq:defI3}) and (\ref{eq:defI3v}) show
that the values of $I_{3u}$ and $I_{3v}$ recovered from initial conditions
are weakly dependent on $\Delta$.

The problem that Fig.~\ref{fig:problem} illustrates is that the blue points
reach the dotted line $p_u^2=0$ significantly to the right of the red curve.
That is, the effective potential defined by the Fudge predicts that the orbit
penetrates to significantly smaller radii than it actually does because $I_3$
is not providing a centrifugal barrier where it should. 

The definition of $I_3$ is such that at a given energy
$I_3$ increases from zero on an orbit that lies within the equatorial plane,
to a maximum value 
\[\label{eq:I3max}
I_{3\rm max}={R^2_{\rm shell}+\Delta^2\over2\Delta^2}v_z^2,
\]
 on the shell orbit $J_r=0$, which moves through the plane at speed   $v_z$.
Hence, when $R_{\rm shell}\gg\Delta$, on a shell orbit
$I_3\simeq\fracj12(J_z/\Delta)^2$
generates a centrifugal exclusion zone around the centre barrier by forcing
$p_u$ to zero at a non-zero value of $u$. 

The black curve in the lower panel of Fig.~\ref{fig:problem} plots the value
of $p_u^2$ that one obtains from equation \eqrf{eq:defI3} with $J_\phi=0$.
This curve, which deviates from the red curve only at small $u$), does {\it
not} dip below the $u$ axis at small $u$, so it fails to exclude the star from
the centre, implying that this is a box orbit; the box/loop transition of
orbits with $J_\phi=0$ occurs at the value of $I_3$ that causes the black
curve's minimum to lie on $p_u^2=0$. In the case of the loop orbit plotted in
Fig.~\ref{fig:problem}, the minimum should lie at $p_u^2<0$.  because $J_z$
is contributing to the orbit's exclusion from the origin.

\subsection{The work-around}\label{sec:solution}

Diminishing the value of $\Delta$ pushes the red and black curves in
Fig.~\ref{fig:problem} down, and for a suitable value of $\Delta$, $p_u^2$ can
be made to turn negative where the blue points terminate, and this is the
best value of $\Delta$. Hence as we move from the shell orbit to more
eccentric loop orbits, we should adjust $\Delta$ such that the black curve
continues to reach the $u$ axis where the blue points terminate. The smallest
value of $u$ at which the black curve can reach $p_u^2=0$ is the location of
the curve's local minimum. The boxes start  when this
minimum first occurs at $p_u^2>0$ so the blue points are free to move to the
centrifugal barrier set by $J_\phi$. For the transition to be modelled as
smooth, as it really is when $|J_\phi|>0$, the location $u_{\rm min}$ of the minimum
as it lifts off the $u$ axis should be close to the location of the
centrifugal barrier set by $J_\phi$ (where the red curves crosses the $u$
axis in Fig.~\ref{fig:problem}).

In summary, the black curve in Fig.~\ref{fig:problem} quantifies the dynamics
in the absence of $J_\phi$. It has a local minimum and if this occurs at $p_u^2>0$,
the star could reach $u=0$ in the absence of $J_\phi$; conversely, when the
minimum occurs at $p_u^2\le0$, the star is excluded from $u=0$ even when
$J_\phi=0$. That is, a centrifugal barrier associated with $J_z$ exists if
and only if the minimum occurs at $p_u^2\le0$. Since exclusion from $u=0$ is
the hallmark of loop orbits, for loop orbits we must ensure that the minimum
occurs at $p_u^2\le0$ while for box orbits the minimum should occur at
$p_u^2>0$. At the box/loop transition, the minimum falls on the $u$ axis.

\begin{figure}
\centerline{\includegraphics[width=\hsize]{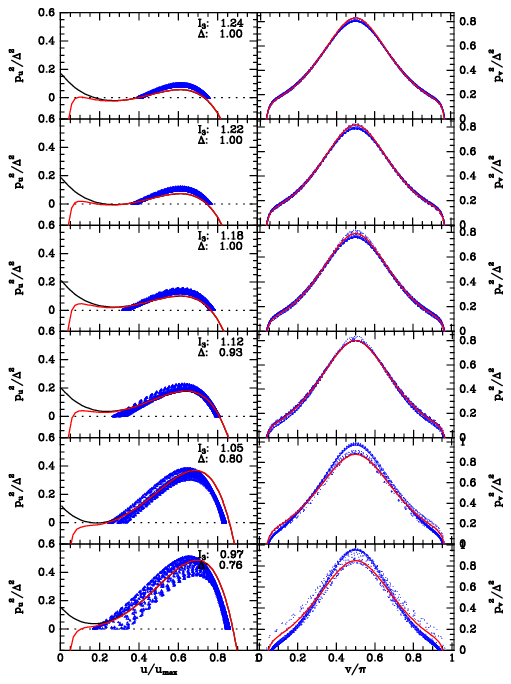}}
\caption{Six orbits plotted in the $(u,p_u^2)$ and $(v,p_v^2)$ planes when
$\Delta$ is given by equation (\ref{eq:interp}). The numbers at top right of the left column are
$I_3/I_{3\rm crit}$ and $\Delta/\Delta_{\rm shell}$. All orbits have $J_\phi$
equal to 5 per cent of the circular angular momentum at $R_{\rm
shell}$.}\label{fig:6vary}
\end{figure}

\begin{figure}
\centerline{\includegraphics[width=\hsize]{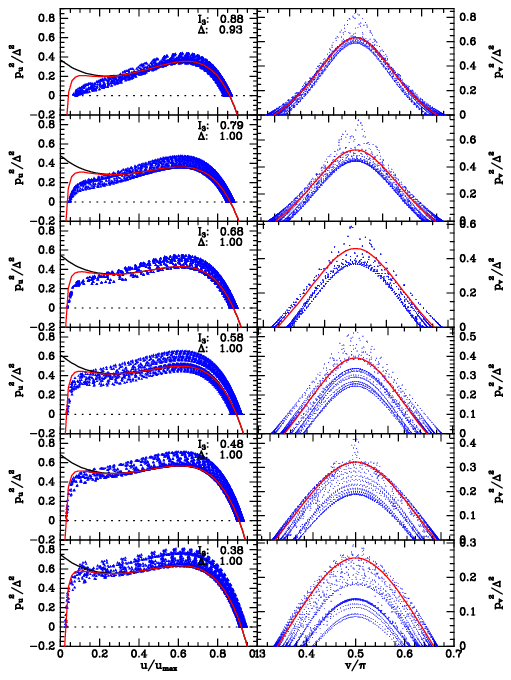}}
\caption{As Fig.~\ref{fig:6vary} but for more eccentric
orbits.}\label{fig:6ecc}
\end{figure}

From the equations $p_u^2=0$ and $\d p_u^2/\d u=0$ we can solve for the
largest focal distance, $\Delta_{\rm excl}$, for which $J_z$ acting alone
excludes the star from $u=0$. During the survey of box/loop transition orbits
that is performed when \agamaTwo\ creates a potential, values of $\Delta_{\rm
crit}(E)=\Delta_{\rm excl}(E,I_{3\rm crit})$ are stored so they can be
subsequently recovered by calling the potential's method {\tt getFDcrit}. The
code also stores the value $u_{\rm min}$ of $u$ at which $p_u^2=0$ -- they
can be recovered from the potential's method {\tt getUmin}.

\begin{figure}
\centerline{\includegraphics[width=.9\hsize]{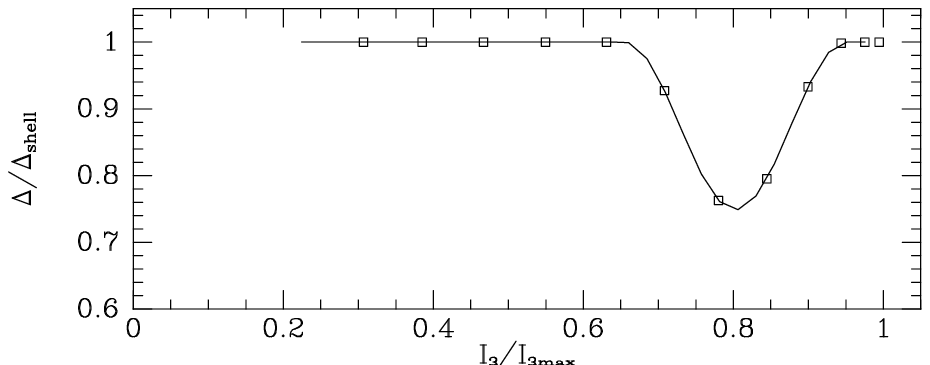}}
\caption{The variation of $\Delta$ with $I_3$ used to produce
Figs.~\ref{fig:6vary} and \ref{fig:6ecc}. The squares correspond to the
orbits in those figures.
}\label{fig:Deltas}
\end{figure}

When action-angle variables are to be computed for {\rd an orbit with
$|J_\phi|<0.1L_{\rm circ}$,}
the orbit's value of $I_3$ is compared with $I_{3\rm crit}$. If $I_3$ is
smaller than $I_{3\rm crit}$, the orbit must be a box and $I_3$ does not
generate a centrifugal barrier. Hence the focal distance must exceed
$\Delta_{\rm excl}(E,I_3)$ --  the code uses the focal distance
$\Delta_{\rm shell}$ obtained from the corresponding shell orbit. If $I_3$ is equal to
$I_{3\rm crit}$, we are dealing with a transition orbit and we should use
$\Delta_{\rm crit}$ so the potential barrier generated by $J_z$ is on the cusp of
disappearing. When $I_3>I_{3\rm crit}$, the focal distance must be such that
$I_3$ creates  a centrifugal barrier, so we need $\Delta<\Delta_{\rm
excl}(E,I_3)$. As $I_3$ tends to its  maximum value, $I_{3\rm max}$ (eqn.~\ref{eq:I3max})
$\Delta$ must tend to $\Delta_{\rm
shell}$. The implemented variation of $\Delta$ with $I_3$ is
\[\label{eq:interp}
\Delta\equiv\begin{cases}
\Delta_{\rm shell}&I_3\le I_{\rm bot}\cr
\overline{\Delta}+\widehat\Delta\cos\psi&
I_{\rm bot}<I_3<I_{\rm top}\cr
\Delta_{\rm shell}&I_3\ge I_{\rm top}
\end{cases},
\]
where
\begin{align}\label{eq:Dbaretc}
I_{\rm bot}&=I_{3\rm crit}-0.75(I_{3\rm max}-I_{3\rm crit})\cr
I_{\rm top}&=I_{3\rm crit}+0.75(I_{3\rm max}-I_{3\rm crit})\cr
\psi&=2\pi{I_3-I_{\rm bot}\over I_{\rm top}-I_{\rm bot}}\cr
\overline{\Delta}\equiv\Delta_{\rm shell}-\widehat\Delta&\quad
\widehat\Delta={\Delta_{\rm shell}-\Delta_{\rm
crit}\over1+\exp\big[4J_\phi^2/(\Delta^2I_{3\rm max})\big]}.
\end{align}
 With this rule, $\Delta$ changes from $\Delta_{\rm shell}$ for $I_3<I_{\rm
bot}$ to $\Delta_{\rm shell}-2\widehat\Delta$ when $I_3=I_{\rm crit}$ and then
grows, reaching $\Delta_{\rm shell}$ at $I_3=I_{\rm top}$. When $J_\phi^2\ll
I_{3\rm max}$, $\Delta_{\rm shell}-2\widehat\Delta\simeq\Delta_{\rm excl}$
but as $J_\phi^2/I_{3\rm max}$ grows, $\widehat\Delta$ diminishes and the
classical Fudge is restored.  Fig.~\ref{fig:Deltas} shows the values of
$\Delta$ that produced Figs.~\ref{fig:6vary} and \ref{fig:6ecc}. 

Figs.~\ref{fig:6vary} and \ref{fig:6ecc} show the $(u,p_u)$ and $(v,p_v)$
planes for twelve orbits of increasing eccentricity when $\Delta$ is given by
equation (\ref{eq:interp}). As in Fig.~\ref{fig:problem}, red and black curves
in the left-hand panels  show the values of $p_u^2$ given by the St\"ackel
formula. They delineate the blue points from the orbit integration reasonably
well, especially when $I_3>I_{3\rm crit}$. Two significant shortcomings
remain:

\begin{itemize}

\item In the third and fourth panels from the top of Fig.~\ref{fig:6vary}
the black curve doesn't quite reach $p_u^2=0$ even though $I_3>I_{3\rm
crit}$. The code deals with this problem by taking $p_u^2=0$ for $u<u_{\rm
min}$, the location of the black curve minimum for $I_3=I_{3\rm crit}$.

\item For the more eccentric orbits the blue points are not nicely
marshalled onto a curve. This failure potentially compromises the returned
values of $J_z$ and $\Omega_z$.

\end{itemize}

\begin{figure}
\centerline{\includegraphics[width=.8\hsize]{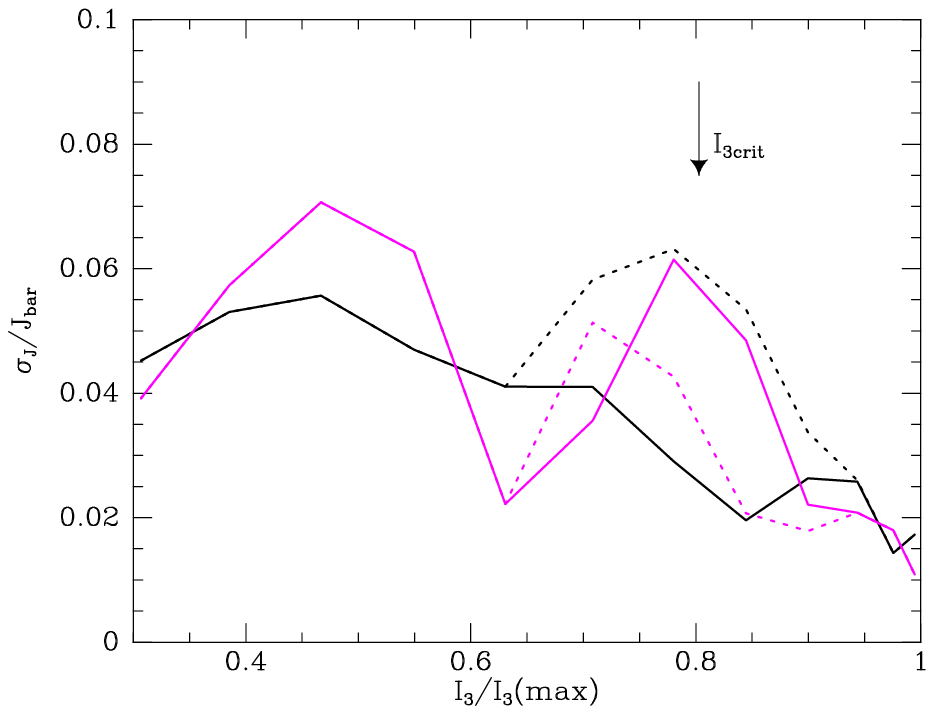}}
 \caption{{\rd Dispersions in actions divided by the mean action
 $(J_r+J_z)/2$ along each of the orbits plotted in
Figs.~\ref{fig:6vary} and \ref{fig:6ecc}. The lines show the standard
deviations of the actions $J_r$ (black) and $J_z$ (magenta) computed at 20
points along the numerically integrated orbit when using $\Delta_{\rm shell}$
(dashed line) or variable $\Delta$ (full line) to eliminate the region of $u$
where $p_u^2<0$.}}\label{fig:acts}
\end{figure}

{\rd The black line in Fig.~\ref{fig:acts} shows the standard deviations in the
values of  $J_r$ returned by the
Fudge along the orbits plotted in Figs.~\ref{fig:6vary} and \ref{fig:6ecc} when using
$\Delta_{\rm shell}$ (dashed line) or the values of $\Delta$ from equation
(\ref{eq:interp}) (full line). The magenta line shows the corresponding
standard deviations in $J_z$.} Using variable $\Delta$ has little impact on the spread in
$J_z$ but it does significantly diminish the spread in $J_r$, especially
when $I_3\simeq I_{3\rm crit}$.

\begin{figure}
\centerline{\includegraphics[width=.7\hsize]{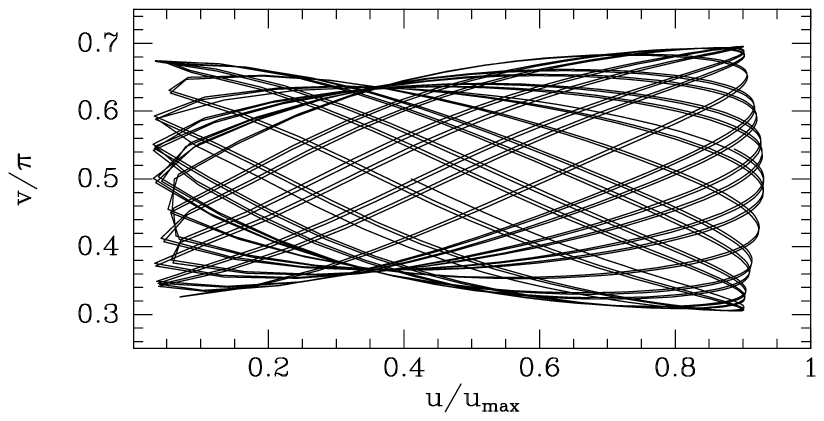}}
\vskip5pt
\centerline{\includegraphics[width=.7\hsize]{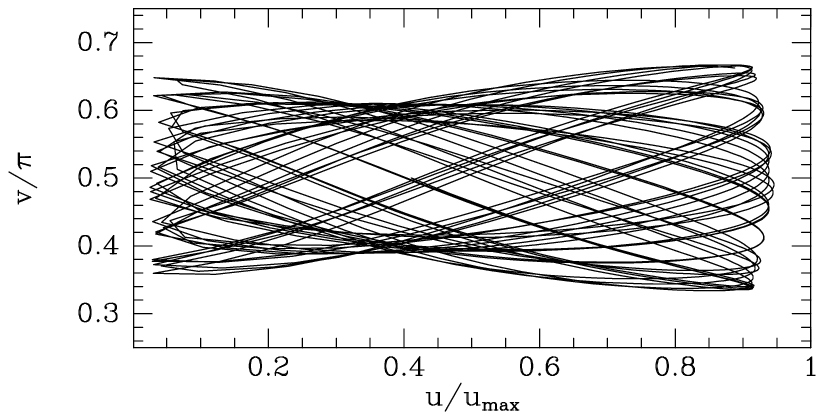}}
\caption{The traces in the $(u,v)$ plane of the orbits giving rise to the
lowest two panels in the column on the extreme  right of
Fig.~\ref{fig:6ecc}. The nearly rectangular trace in the upper panel here
is associated with a single arc in the $(v,p_v^2)$ plane.}\label{fig:grid1}
\end{figure}

Fig.~\ref{fig:grid1} gives insight into why the blue points in the bottom
right panel of Fig.~\ref{fig:6ecc} are grouped into two distinct arcs. It
shows the $(u,v)$ traces of that orbit (lower panel) and the trace of the
orbit above it in Fig.~\ref{fig:6ecc}, which produces a single (wide) arc in
the $(v,p_v^2)$ plane. In the lower panel of Fig.~\ref{fig:grid1}, the orbit
has three distinct types of turning point in $v$. Two of these occur at the
left and right corners of the roughly rectangular trace of the orbit. These
turning points are associated with the feet of the larger arc in the bottom
right panel of Fig.~\ref{fig:6ecc}.  The third type of turning point occurs
at $u\simeq0.35u_{\rm max}$, $v\simeq0.61\pi$ and is associated with the feet
of the smaller arc in Fig.~\ref{fig:6ecc}. The turning points of the orbit
plotted in the upper panel of Fig.~\ref{fig:grid1} are all of the first or
second type, so the orbit generates a single arc in the $(v,p_v^2)$ plane.

Near the middle of the upper edges of the traces in both panels there are
regions in which the star transits along three lines rather than two.
The existence of these regions implies that the orbit's third
integral cannot be expressed as a quadratic in $p_u$ and $p_v$ as the Fudge requires.
Fortunately  the lines fall into two groups of nearly parallel lines so the
Fudge can provide good approximations to the true velocities.

\subsection{Frequencies}\label{sec:freq}

\begin{figure}
\centerline{\includegraphics[width=.8\hsize]{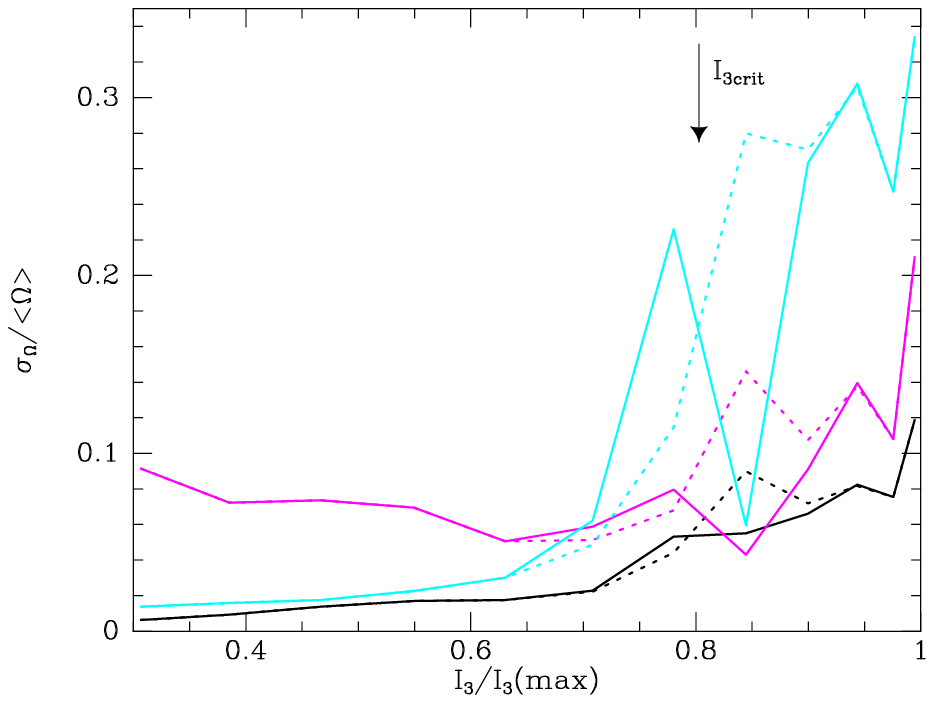}}
 \caption{{\rd Dispersions in frequencies normalised by the corresponding
 frequency along each of the orbits plotted in
Figs.~\ref{fig:6vary} and \ref{fig:6ecc}. The lines show the standard
deviations in the computed frequencies $\Omega_r$ (black), $\Omega_z$ (magenta) and
$\Omega_\phi$ (cyan) computed at 20
points along the numerically integrated orbit when using $\Delta_{\rm shell}$
(dashed line) or variable $\Delta$ (full line) to eliminate the region of $u$
where $p_u^2<0$.}}\label{fig:freqs}
\end{figure}

Orbital frequencies sometimes play a significant role in galactic dynamics,
so we consider improvements in how \agamaTwo\ computes them. 

The Fudge computes frequencies  by inverting the matrix $\p J_i/\p I_j$, where
$\vI=(E,I_3,J_\phi)$.\footnote{The matrix $\p J_i/\p I_j$ is also involved in
the computation of the angle variables $\theta_i(\vx,\vv)$.} When the
integral for $J_i$ is differentiated, $p_i$ migrates to the denominator of
the integrand.  For example
\[
{\p J_r\over\p I_3}=-{1\over\pi}\int_{u_{\rm min}}^{u_{\rm max}}{\d u\over
p_u}.
\]
 With the setup illustrated by Fig.~\ref{fig:problem}, the adopted value of
$u_{\rm min}$ is too small because it is where the red curve passes through
zero rather than the endpoint of the blue points, and $p_u$ may be predicted
to be very small in the spurious interval of integration. Hence the
contribution of the spurious interval to $\p J_r/\p I_3$ can then be much
greater than its contribution to $J_r\propto\oint\d u\,p_u$. It follows that
the Fudge is liable to compute frequencies markedly less accurately than it
computes actions.

{\rd Fig.~\ref{fig:freqs} shows standard deviations in the computed frequencies of the orbits
plotted in Figs.~\ref{fig:6vary} and \ref{fig:6ecc} when $\Delta$ is set to
$\Delta_{\rm shell}$ (dashed line) and when it is varied according to
equation (\ref{eq:interp}) (full line).} With fixed $\Delta$
serious errors are apparent for $I_3>I_{3\rm crit}$. Varying $\Delta$ 
modestly reduces the error in $\Omega_r$ and $\Omega_z$, especially near
$I_{3\rm crit}$ but does not provide
a satisfactory resolution of the problem. The errors in $\Omega_\phi$ remain
unacceptably large. {\rd This weakness of the Fudge should be borne in mind on the
comparatively rare occasions when it's used to compute frequencies rather
than actions.}

\section{Conclusions}\label{sec:conclude}

Even at non-zero $J_\phi$ it is useful to divide orbits in an axisymmetric
potential into box and loop orbits. $J_z$ contributes to the
centrifugal barrier experienced by orbits only in the case of loop orbits. A
procedure was described for computing the critical action $\Jcrit$ below
which orbits are boxes. Knowledge of $\Jcrit$ proves to be important when
constructing the tori of eccentric orbits \citep{BinneyVasilievWright} and
when evaluating the distribution function of a spheroidal component
\citep{BinneyDFs2025}.

Particular care in the choice of the focal distance $\Delta$ is required when
applying the St\"ackel Fudge to orbits that have $J_z\simeq \Jcrit$.  The
optimum value of $\Delta$ is the one that correctly predicts the smallest
value of the ellipsoidal variable $u$ to which the orbit penetrates. In an
interval around $\Jcrit$, this value proves smaller than the value
$\Delta_{\rm shell}$ that fits the shell orbit. The St\"ackel Fudge action
finder provided by \agamaTwo\ reduces $\Delta$ in the vicinity of $\Jcrit$,
and as a consequence significantly reduces the scatter in the values of $J_r$
computed along eccentric orbits. The computational cost of this upgrade is
negligible.

Values of orbital frequencies are particularly vulnerable to sub-optimal
choices of $\Delta$. Unfortunately, the change that reduces the scatter in
$J_r$ has less impact on the scatter in frequencies. 

Understanding that orbits in the meridional plane are of two distinct types
proves crucial (a) for torus mapping, and (b) for structuring distribution
functions $f(\vJ)$ for stellar systems with anisotropic velocity
distributions.  \cite{BinneyVasilievWright} extend torus mapping to almost
radial orbits in flattened potentials by using harmonic-oscillator or
isochrone toy maps according as $J_z$ is smaller or larger than $\Jcrit$.
Possession of a convenient way of computing $\Jcrit(J_{\rm fast})$ enabled
\cite{BinneyDFs2025} to construct the first action-based DFs for anisotropic,
flattened systems that yield physical velocity distributions in the
neighbourhood of $v_\phi=0$.  Consequently, we expect the algorithm for
computing $\Jcrit(J_{\rm fast})$ given in Section \ref{sec:getJcrit} and embedded in
\agamaTwo\ to play a significant role as angle-action variables are used to
interpret data for galaxies of all types.

\section*{Acknowledgements}

It's a pleasure to thank Eugene Vasiliev for extensive assistance in
understanding why {\sc agama}'s Fudge was performing badly.
This research was supported by the Leverhulme Trust under grant LIP-2020-014
and in part by grant NSF PHY-2309135 to the Kavli
Institute for Theoretical Physics (KITP). 

\section*{Data Availability}

The computations were performed by C++ code linked to the \agamaTwo\ library
\citep{BinneyVasilievWright}, which can be downloaded from {\tt
https://github.com/binneyox/AGAMAb}.

\bibliographystyle{mn2e} \bibliography{/u/tex/papers/mcmillan/torus/new_refs}

\begin{thebibliography}{18}
\expandafter\ifx\csname natexlab\endcsname\relax\def\natexlab#1{#1}\fi

\bibitem[{{Binney}(2012{\natexlab{a}})}]{JJB12:Stackel}
{Binney} J., 2012{\natexlab{a}}, \mnras, 426, 1324

\bibitem[{{Binney}(2012{\natexlab{b}})}]{JJB12:dfs}
{Binney} J., 2012{\natexlab{b}}, \mnras, 426, 1328

\bibitem[{{Binney}(2026)}]{BinneyDFs2025}
{Binney} J., 2026, arXiv, 2602.23127

\bibitem[{{Binney} \& {McMillan}(2016)}]{JJBPJM16}
{Binney} J., {McMillan} P.~J., 2016, \mnras, 456, 1982

\bibitem[{{Binney} \& {Sch{\"o}nrich}(2018)}]{BinneySchoenrich2018}
{Binney} J., {Sch{\"o}nrich} R., 2018, \mnras, 481, 1501

\bibitem[{{Binney} \& {Tremaine}(2008)}]{GDII}
{Binney} J., {Tremaine} S., 2008, {Galactic Dynamics: Second Edition}.
  Princeton University Press

\bibitem[{{Binney} {et~al}\mbox{.}(2025){Binney}, {Vasiliev}, \&
  {Wright}}]{BinneyVasilievWright}
{Binney} J., {Vasiliev} E., {Wright} T., 2025, arXiv, 2512.06512

\bibitem[{{Binney} \& {Wong}(2017)}]{BinneyWong}
{Binney} J., {Wong} L.~K., 2017, \mnras, 467, 2446

\bibitem[{{Bland-Hawthorn} {et~al}\mbox{.}(2019){Bland-Hawthorn}, {Sharma},
  {Tepper-Garcia}, {Binney}, {Freeman}, {Hayden}, {Kos}, {De Silva}, {Ellis},
  {Lewis}, {Asplund}, \& {Buder}}]{JBH_Galah2019}
{Bland-Hawthorn} J. {et~al.}, 2019, \mnras, 486, 1167

\bibitem[{{Bland-Hawthorn} \& {Tepper-Garcia}(2021)}]{JBHThor2020}
{Bland-Hawthorn} J., {Tepper-Garcia} T., 2021, \mnras, 504, 3168

\bibitem[{{Bovy}(2014)}]{Bovy2014:streams}
{Bovy} J., 2014, \apj, 795, 95

\bibitem[{{Bovy} \& {Rix}(2013)}]{Bovy2013}
{Bovy} J., {Rix} H.-W., 2013, \apj, 779, 115

\bibitem[{{de Zeeuw}(1985)}]{deZeeuw1985}
{de Zeeuw} T., 1985, \mnras, 216, 273

\bibitem[{{Navarro} {et~al}\mbox{.}(1997){Navarro}, {Frenk}, \&
  {White}}]{NFW97}
{Navarro} J.~F., {Frenk} C.~S., {White} S.~D.~M., 1997, \apj, 490, 493

\bibitem[{{Sanders} \& {Binney}(2013)}]{SaJJB13b}
{Sanders} J.~L., {Binney} J., 2013, \mnras, 433, 1826

\bibitem[{{Sanders} \& {Binney}(2014)}]{SaJJB14}
{Sanders} J.~L., {Binney} J., 2014, \mnras, 441, 3284

\bibitem[{{Vasiliev}(2019{\natexlab{a}})}]{AGAMA}
{Vasiliev} E., 2019{\natexlab{a}}, \mnras, 482, 1525

\bibitem[{{Vasiliev}(2019{\natexlab{b}})}]{Vasiliev2019}
{Vasiliev} E., 2019{\natexlab{b}}, \mnras, 484, 2832

\end{thebibliography}
\end{document}